\documentclass[12pt]{article} 
\usepackage{epsf}

\begin{document}
\begin{center}
{\bf \Large A New Property of the Quark-Antiquark Potential in a Quark-Gluon Plasma}
\end{center}
\vspace*{1cm}
\begin{center}
{\bf Sidi Cherkawi Benzahra\footnote[1]{benzahra@physics.spa.umn.edu}}
\end{center}
\begin{center}
{\large School of Physics and Astronomy}
\end{center}
\begin{center}
{\large University of Minnesota}
\end{center}
\begin{center}
{\large Minneapolis, MN 55455}
\end{center}
\vspace*{1cm}
\begin{abstract}
\noindent

I consider the behavior of the quark-antiquark potential, called the Cornell potential, in a quark-gluon plasma. Since 
mesons are no longer bound in the quark-gloun plasma, there might be a relationship between the string tension of the 
quark-antiquark potential, the mass of the quark, and the coupling constant of the meson.   
\end{abstract}

\hspace{0.3in} When the quark or anti-quark is struck by a high energy gluon, the meson can dissociate into other elements.  
The medium, the quark-gluon plasma, can be full of gluons that can cause this dissociation, and this can happen by exciting 
color-singlet state ${\mid Q\bar{Q} \rangle}^{(1)}$ into a color-octet continuum state, ${\mid Q\bar{Q} \rangle}^{(8)}$. 
The quark absorbs energy from the gluon field.  In addition to the dissociation of mesons by gluons, there is another kind 
of dissociation which is caused by the screening of the color charges of the quarks in the medium [1].  In the 
high temperature deconfined phase, the quark-antiquark free energy $V_{Q\bar{Q}}$, which is the Debye potential with 
inverse screening length, $m_{el}$, is given by [2]

\begin{equation}
{V_{Q\bar{Q}}}=-{4 \over {3}}{{\alpha_{s}} \over {r}}e^{-m_{el}r} \, .
\end{equation}

Using a variational calculation with an exponential trial wavefunction,
$Ae^{-r/a_{T}}$, I looked for a critical value of $m_{el}$ where the
upsilon meson is no longer bound. Here, $a_{T}$ is just a parameter.  I found this critical $m_{el}$ to be [3]
\begin{equation}
{m_{el}}= {2 \over {3}} \alpha_{s} m_{Q} \, .
\end{equation}

Here I used [4] $\alpha_{s}(m_{b}) = 0.2325 \pm 0.0044$.  Inserting $m_{el}$ 
in the equation [2]
\begin{equation}
{m_{el}^{2}} = {1 \over {3}} g^{2} (N + {{N_{f}} \over {2}}) T^{2} \, ,
\end{equation}
where N=3 from the SU(N) group and $N_{f}=3$ is the number of light flavors, 
and using
the temperature-dependent coupling constant as given by [2]
\begin{equation}
{{g^{2}} \over {4 \pi}} = {{6 \pi} \over {27 \rm ln \left(T/50 \rm MeV \right)}}
\end{equation}

I found that the ground state of upsilon is unbound at a temperature of 
T=250 MeV. Above this temperature, the effect of screening does not allow 
the upsilon meson to exist in a 1s state.\\

The quark-antiquark potential
\begin{equation}
V=-{{4}\over{3}}{{\alpha_{s}}\over {r}} + b r \, ,
\end{equation} 
will be treated the same way.  This potential is just a model. Since quarks have not
been experimentally observed, it was plausible to postulate a potential which is of a coulomb type at short distances--at 
extremely short distances $b$ and $\alpha_{s}$ decrease, leading to asymptotic freedom--and grows linearly at greater
separations, thus leading to confinement of quarks in hadrons.  But in quark-gluon plasma, this potential will not behave
the same.  I will use a variational calculation with the same trial wave function, $Ae^{-r/a_{T}}$, and 
look for what value the string tension, b, in the quark-antiquark potential might have, when the meson is no longer bound.  
The binding energy of the meson is:

\begin{equation}
{E} = {{{1} \over {{ m_{ Q} a_{ T}^{ 2}}}} - {{ 4} \over { 3}} {{\alpha_{ s}} \over {a_{ T}}}} + {{3} \over {2}} b a_{T}\, .
\end{equation}   

The binding energy in terms of $a_{T}$ has only one minimum. Taking the derivative of E with respect to parameter $a_{T}$ 
and making it vanish in order to be at that minimum, gives me another equation in terms of b and $a_{T}$:
\begin{equation}
{-{{2} \over {{ m_{ Q} a_{ T}^{ 3}}}} + {{ 4} \over { 3}} {{\alpha} \over {a_{ T}^{2}}}} + {{3} 
\over {2}} b = 0 \, .
\end{equation}

Since the meson is no longer bound in the quark-gluon plasma, I set E equal to zero, and this gives me a value for $a_{T}$
and a value for b, which are

\begin{equation}
a_{T}= {9 \over 8\alpha_{s}m_{Q}} \quad ; \quad b={1\over 3}\left ({8 \over 9}\right )^3 \alpha_{s}^{3} m_{Q}^{2} \, .
\end{equation} 

Since b is not supposed to depend on the mass of the quark, but here, as indicated in eq.(6) depends on it, this proves that
the Cornell potential no longer holds in the quark-gluon plasma. 

\begin{center}
{\bf \large Conclusion}
\end{center}

The string tension, b, was 
previously determined by fitting the principal energy levels of the quark anti-quark states from the nonrelativistic 
Schrodinger equation with the Cornell potential.  Typical result: $b \approx 1 \rm GeV/fm$.  But here, I find it to take 
the values of 0.267 GeV/fm for $\Upsilon$ and 0.054 Gev/fm for $J/\psi$.  Finally, Since the Cornell potential breaks down 
in quark-gluon plasma, it is not a consistent potential.

\begin{center}
{\bf \large References}
\end{center}
[1] T. Matsui and H. Satz, Physics Letters B 178, 416 (1986). \newline
[2] J. I. Kapusta, Finite-Temperature Field Theory (Cambridge University 
Press, Cambridge, 1989). \newline
[3] S. C. Benzahra Phys. Rev. C 61, 064906 (2000) \newline
[4]Matthias Jamin and Antonio Pich, Nucl. Phys. B507, 334 (1997).\newline

\end{document}